\documentclass[a4paper]{article}
\usepackage[english]{babel}
\input{epsf}
\usepackage{mathtext}
\usepackage{graphics,dcolumn,bm,float}
\usepackage{amssymb,amsmath,rotate,color}
\usepackage{times}
\usepackage{color}
\usepackage{fancyhdr}
\newcommand{\be}{\begin{equation}}
\newcommand{\ene}{\end{equation}}
\newcommand{\ba}{\begin{array}}
\newcommand{\ea}{\end{array}}

\begin{document}
\title{Fuzzy AdS-conifold $ Y_{AdS_{F}}^{6} $ and Dirac operator of principal fibration $ X_{AdS_{F}}^{5}\rightarrow AdS_{F}^{2}\times AdS_{F}^{2} $}
\author{M. Lotfizadeh\thanks{E-mail: M. Lotfizadeh@urmia.ac.ir}$^{\;\;,1}$  \\
\footnotesize\textit{$^1$Department of Physics, Faculty of Science, Urmia University, P.O.Box: 165, Urmia, Iran}\\
\footnotesize\textit{}}
% Remove command to get current date 
\date{}
\maketitle
\begin{abstract}
It has been constructed fuzzy AdS- conifold $ Y_{AdS_{F}}^{6} $ on the base $ AdS_{F}^{3}\times AdS_{F}^{2} $ which is topologically homeomorphic with the total space of the fibration $ X_{AdS_{F}}^{5}\rightarrow AdS_{F}^{2}\times AdS_{F}^{2} $. After the projective module description of this bundle, the pseudo fuzzy Dirac and chirality operators on fuzzy $ AdS^{2}_{F}\times AdS_{F}^{2} $ have been studied. Using the fuzzy Ginsparg-Wilson algebra, it has been studied the gauged fuzzy Dirac and chirality operators in instanton sector. It has been showed that they have correct commutative limit in the limit case when noncommutative parameter $ l_{\alpha} $ tends to infinity.
\end{abstract}
\textbf{PACS}: 74.45.+c; 85.75.-d; 73.20.-r\\
\textbf{Keywords}: fuzzy AdS-conifold, pseudo gauged Ginsparg-Wilson algebra, pseudo gauged Dirac and chirality operators.

\section{Introduction}
Conifolds are singular complex spaces that are $ C^{\infty} $ except in a number of isolated conical singularities. In the neighbourhood of these singularities the conifolds are described by 
\begin{equation*}
\sum_{\alpha=1}^{n}z_{\alpha}^{2}=0.
\end{equation*}
We consider the singular point in the origin of $ \mathbb{C}^{n} $. It has been showed that the first Chern class of these spaces are zero [1].The $ (2n-2)- $dimensional conifold $ Y^{2n-2} $ is a noncompact Calabi-Yau manifold with $ SO(n)\times U(1) $ as its symmetry group. The base manifold of the conifold $ Y^{6} $ is a $ (2n-3)- $dimensional manifold $ X^{2n-3} $ which is a compact Einstein manifold and satisfies $ R_{ij}= (2n-4)g_{ij} $. The manifold $ X^{2n-3} $ is the intersection of $ Y^{2n-2} $ with the sphere $ S^{2n-1} $. The conifold $ Y^{2n-2} $ is a cone over $ X^{2n-3} $. The $ n=3 $ fuzzy case is the four-dimensional Fuzzy conifold $ Y_{F}^{4} $ with $ X_{F}^{3} $ as its base. The manifold $ X_{F}^{3} $ is the intersection of $ Y_{F}^{4} $ with the sphere $ S_{F}^{5} $: $ X_{F}^{3}= Y_{F}^{4}\cap S_{F}^{5} $. The fuzzy version of these spaces and also the $ U(1) $ principal monopole bundle  $ X_{F}^{3}\rightarrow S_{F}^{2} $ has been studied in [2]. Another important case, is the case $ n=6 $. The fuzzy compact manifold $ X_{F}^{8} $ is the intersection of the fuzzy ten-dimensional conifold $ Y_{F}^{10} $ with $ S_{F}^{5}\times S_{F}^{5} $. $ X_{F}^{8} $ is the base of $ Y_{F}^{10} $. Authors in [3] have studied monopoles and Dirac operator of the principal fibration $ X_{F}^{8}\rightarrow S_{F}^{5}\times S_{F}^{5} $. The most important case is the case $ n=4 $ [4-7]. The fuzzy compact manifold $ X_{F}^{5} $ is the intersection of the fuzzy six-dimensional conifold $ Y_{F}^{6} $ with $ S_{F}^{2}\times S_{F}^{2} $. The manifold $ X_{F}^{5} $ is the base of $ Y_{F}^{6} $. In [4] it has been constructed fuzzy conifold $ Y_{F}^{6} $ and its fuzzy base $ X_{F}^{5} $. Also, it has been showed that $ X_{F}^{5} $ is a $ U(1) $ principal fibration over fuzzy $ S^{2}\times S^{2} $. In [5], $ U(n) $ gauge theory on fuzzy $ S^{2}\times S^{2} $ as a matrix model has been studied. Authors have showed that this gauge theory reduces to Y-M gauge theory on $ S^{2}\times S^{2} $ in commutative limit. Quantum effective potential for $ U(1) $ fields on fuzzy $ S^{2}\times S^{2} $ has been studied in [6]. In [7], it has been studied the construction of a topological charge on fuzzy $ S^{2}\times S^{2} $ via a Ginsparg-Wilson relation. In present paper, it has been constructed fuzzy AdS- conifold $ Y_{AdS_{F}}^{6} $ on the base $ AdS_{F}^{3}\times AdS_{F}^{2} $ which is topologically homeomorphic with the total space of the fibration $ X_{AdS_{F}}^{5}\rightarrow AdS_{F}^{2}\times AdS_{F}^{2} $.

Dirac and chirality operators are two important self-adjoint operators for the Connes-Lott approach to noncommutative geometry[8,9]. There are three types of Dirac and chirality operators. Ginsparg-Wilson Dirac operator, $ D_{GW} $ [10-18], Watamura-Watamura Dirac operator $ D_{WW} $ [19-21] and Grosse-Klimcik-Presnajder Dirac operator $ D_{GKP} $ [22, 23]. These three types of Dirac operators are compared with each other in [24].

 In this paper we generalize Ginsparg-Wilson algebra to  pseudo fuzzy Ginsparg-Wilson algebra on $ AdS_{F}^{2}\times AdS_{F}^{2} $ and construct its pseudo Dirac and chirality operators. This paper is organized as follows: In section 2 we study the fuzzy AdS-conifold $ Y_{AdS_{F}}^{6} $ and fibre bundle structure of the fibration $ X_{AdS_{F}}^{5}\rightarrow AdS_{F}^{2}\times AdS_{F}^{2}$. Projective module description of the fibration $ X_{AdS_{F}}^{5}\rightarrow AdS_{F}^{2}\times AdS_{F}^{2} $ has been studied in section 3. In section 4, spin $\frac{1}{2}$ projectors of the pseudo projective $ A(AdS_{F}^{2}\times AdS_{F}^{2}) -$module has been constructed.
Fuzzy Ginsparg-Wilson algebra and its Spin $\frac{1}{2}$ fuzzy pseudo Dirac and chirality operators on $ AdS_{F}^{2}\times AdS_{F}^{2} $ have been studied in section 5. In section 6,  gauged pseudo Dirac and chirality operators was constructed. In section 7, instanton coupling and in section 8, gauging the pseudo fuzzy Dirac operator in instanton sector have been studied, respectively.

\section{Fuzzy AdS-conifold $ Y_{AdS_{F}}^{6} $ and fibre bundle structure of the fibration $ X_{AdS_{F}}^{5}\rightarrow AdS_{F}^{2}\times AdS_{F}^{2} $}

The AdS-conifold $ Y_{AdS}^{6} $ is a six-dimensional manifold embedded in a four-dimensional complex space $ \mathbb{C}^{4}$ with four complex coordinates $ w_{i}(i= 1,...,4) $ satisfying:
\begin{equation}
w_{1}^{2}+w_{2}^{2}-w_{3}^{2}-w_{4}^{2}= w_{i}\eta^{ij}w_{j}=0, \quad i,j=1,2,3,4,\tag{2-1}
\end{equation}
where $ \eta_{ij}= diag(+1,+1,-1,-1) $ and we have used the Einstein sumation convention. 
From (2-1) one can calculate the base of the $ Y_{AdS}^{6} $ by intersecting the space of solutions of (2-1) with the manifold $ AdS^{7} $ of radius $ r $ in $ \mathbb{C}^{4} $
\begin{equation}
|w_{1}|^{2}+ |w_{2}|^{2}-|w_{3}|^{2}-|w_{4}|^{2}= w_{i}\eta^{ij}\bar{w_{j}}=-r^{2}.\tag{2-2}
\end{equation}
If we break up $ w_{i} $ into its real and imaginary parts, $ w_{i}= x_{i}+y_{i} $ then it is easy to see that from (2-1) and (2-2) we have:
\begin{equation*}
x_{i}\eta^{ij}x_{j}= x_{1}^{2}+x_{2}^{2}-x_{3}^{2}-x_{4}^{2}= -\dfrac{r^{2}}{2•}, 
\end{equation*}
\begin{equation}
 y_{i}\eta^{ij}y_{j}= y_{1}^{2}+y_{2}^{2}-y_{3}^{2}-y_{4}^{2}= -\dfrac{r^{2}}{2•}, \quad x_{i}\eta^{ij}y_{j}= x_{1}y_{1}+x_{2}y_{2}-x_{3}y_{3}-x_{4}y_{4}=0.\tag{2-3}
\end{equation}
The first of these equations defines an $ AdS^{3} $ with radius $ r/\sqrt{2}$. The other two equations define an $ AdS^{2} $ principal $ U(1) $ fibration over $ AdS^{3} $ 
\begin{equation}
U(1)\hookrightarrow AdS^{3}\rightarrow AdS^{2}.\tag{2-4}
\end{equation}
Because all such bundles are trivial, the base of $ Y_{AdS}^{6} $ has a topology of $ AdS^{2}\times AdS^{3} $.
The $ 3- $dimensional anti de Sitter space $ AdS^{3} $ is a maximally symmetric space with constant negative curvature. It is the hyperboloid $ AdS^{3}\hookrightarrow \mathbb{R}^{2,2} $. The isometry group of $ AdS^{3} $ is $ SO(2,2)\simeq SU(1,1)\times SU(1,1)\simeq AdS^{2}\times AdS^{2} $ with the Lie algebra $ so(2,2)\simeq su(1,1)\oplus su(1,1) $.
 The base of the $ Y_{AdS}^{6} $ is $ X_{AdS}^{5} $ which is the intersection of $ Y_{AdS}^{6} $ with the hyperboloid $ AdS^{7}(\bar{z_{i}}\eta^{ij}z_{j}= -r^{2}) $ i.e. $ X_{AdS}^{5}= Y_{AdS}^{6}\bigcap AdS^{7} $ for $ r= cons > 0 $, the manifold $ X_{AdS}^{5} $ is a differentiable five-dimensional manifold.
The conifold $ Y_{AdS}^{6} $ has a class of manifolds $ T_{AdS}^{p,q} $ (p and q are integers) as its base. These $ T_{AdS}^{p,q ,} $s are topologically homeomorphic with the manifold $ AdS^{3} \times AdS^{2} $, but they are not geometrically equivalent i.e. they are not diffeomorphic. All of $ T_{AdS}^{p,q}, $s are $ U(1) $ principal bundles  over the cross manifold $ AdS^{2}\times AdS^{2} $. The manifold $ AdS^{2}\times AdS^{2} $ can be defined by the coordinates $ x_{i_{1}} $ and $ x_{i_{2}} $ as:
\begin{align*}
x_{i_{1}}= x_{i_{1}}(\theta_{1}, \xi_{1}), \quad x_{i_{1}}\eta^{i_{1}j_{1}}x_{j_{1}}= -r_{1}^{2},
\end{align*} 
\begin{equation}
x_{i_{2}}= x_{i_{2}}(\theta_{2}, \xi_{2}), \quad x_{i_{2}}\eta^{i_{2}j_{2}}x_{j_{2}}= -r_{2}^{2},\quad i_{1,2}=1,2,3.\tag{2-5}
\end{equation}
As $ SU(1,1) $ is isomorphic with $ AdS^{3} $ so we have:
\begin{equation}
AdS^{3} \times AdS^{2}= SU(1,1)\times \frac{SU(1,1)}{U(1)•}.\tag{2-6}
\end{equation}
For the special case when $ p= q= 1 $ we have $ X_{AdS}^{5}(:= T^{1,1}) $. For the manifold $ X_{AdS}^{5} $ it is not important the $ U(1) $ is quotiented from which $ SU(1,1) $. So one can write:
\begin{equation}
X_{AdS}^{5} \cong \frac{SU(1,1)\times SU(1,1)}{U(1)•}.\tag{2-7}
\end{equation}
It is clear that both $ X_{AdS}^{5}(=T_{AdS}^{1,1}) $ and $ AdS^{3}\times AdS^{2}(:=T_{AdS}^{1,0}) $ are principal $ U(1) $ bundles over $ AdS^{2}\times AdS^{2} $. As $ Y_{AdS}^{6} $ has the $ SO(2,2)\times U(1) $ symmetry, the symmetry group of $ X_{AdS}^{5} $ is $ SO(2,2) $. $ SO(2,2) $ is a Lie group generated by $ \lbrace M_{1}, M_{2}, M_{3}, T_{1}, T_{2}, T_{3}\rbrace $ satisfying:
\begin{equation}
[M_{i}, M_{j}]= iC_{ij}^{\;\;\;k}M_{k}, \quad [M_{i}, T_{j}]= iC_{ij}^{\;\;\;k}T_{k}, \quad [T_{i}, T_{j}]= iC_{ij}^{\;\;\;k}M_{k},  \quad i,j,k =1,2,3.\tag{2-8}
\end{equation}
where $ C_{ij}^{\;\;\;k} $  are determined as $ C_{ij}^{\;\;\;k}=\eta^{kl} C_{ijl} $, in which $ C_{123}=1 $ and $ C_{ij}^{\;\;\;k} $ are completely antisymmetric. The Minkowskian metric $ \eta^{ij} = \eta_{ij} = diag(1,1,-1,-1) $ raises and lowers the indexes. The structure constants $C_{ij}^{\;\;\;k}$ satisfy the following relation:
\begin{equation}
C_{im}^{\quad k} \eta^{ij}C_{jl}^{\;\;\;n} = \eta_{m}^{n} \eta_{l}^{k} -\eta_{ml} \eta^{kn}.
\tag{2-9}
\end{equation}
The generators of $su(1,1)$ Lie algebra are pseudo Hermitian with respect to $\Lambda $:
\begin{equation}
M_{i}^{\dagger} = \Lambda M_{i}\Lambda^{-1}, \quad T_{i}= \Lambda T_{i}\Lambda^{-1},
\tag{2-10}
\end{equation}
where the operator $ \Lambda $ satisfies $ \Lambda^{\dagger}=\Lambda, \Lambda^{2}=1 $ and $ \Lambda^{-1}=\Lambda^{\dagger}. $

Let us define the new operators $ Q_{i} $ and $ \bar{Q_{i}} $ as:
\begin{equation}
Q_{i}= \frac{1}{2•}(M_{i}+T_{i}), \quad \bar{Q_{i}}= \frac{1}{2•}(M_{i}-T_{i}).\tag{2-11}
\end{equation}
Using the new operators, the Lie algebra $ so(2,2) $ decomposes into two disjoint $ su(1,1) $ algebra:
\begin{equation}
[Q_{i}, Q_{j}]= iC_{ij}^{\;\;\;k}Q_{k}, \quad [\bar{Q_{i}}, \bar{Q_{j}}]= iC_{ij}^{\;\;\;k}\bar{Q_{k}},\quad  \quad [Q_{i}, \bar{Q_{j}}]= 0.\tag{2-12}
\end{equation}
To see the relation of $ SO(2,2) $ with two $ SU(1,1)^{,} $s, let us define:
\begin{equation}
A_{i}= U^{\dagger}Q_{i}U= \frac{1}{2•}I_{2}\otimes \Sigma_{i}, \quad B_{i}= U^{\dagger}\bar{Q_{i}}U= \frac{1}{2•}\Sigma_{i}\otimes I_{2},\tag{2-13}
\end{equation}
where
\begin{equation}
U= \frac{1}{\sqrt{2}•}\begin{pmatrix}
1  &0  &0  &-1\\
i  &0  &0  &i\\
0  &-1  &-1  &0\\
0  &-i  &i  &0
\end{pmatrix},\tag{2-14}
\end{equation}
and $ \Sigma_{1}=i\sigma_{1}, \Sigma_{2}=i\sigma_{2} $ and $ \Sigma_{3}=\sigma_{3} $ ($ \sigma_{i} $ are Pauli matrices.) are the generators of $ su(1,1) $ Lie algebra
\begin{equation}
[\Sigma_{i}, \Sigma_{j}]= iC_{ij}^{\;\;\;k}\Sigma_{k}.\tag{2-15}
\end{equation}

One can easily check that:
\begin{equation}
[A_{i}, A_{j}]= iC_{ij}^{\;\;\;k}A_{k}, \quad [B_{i}, B_{j}]= iC_{ij}^{\;\;\;k}B_{k}, \quad [A_{i}, B_{j}]= 0.\tag{2-16}
\end{equation}
Now let us define the fibre projection map $ \pi: X_{AdS}^{5}\rightarrow AdS^{2}\times AdS^{2} $ as:
\begin{equation}
x_{i_{1}}= z^{\dagger}A_{i_{1}}z, \quad x_{i_{2}}= z^{\dagger}B_{i_{2}}z, \quad i=1,2,3,\tag{2-17}
\end{equation}
with $ z= (z_{1}, z_{2}, z_{3}, z_{4})^{t} \in \mathbb{C}^{4} $. It is obvious that $ x_{i_{1}} $ and $ x_{i_{2}} $ are real and satisfies the following relation:
\begin{equation}
x_{i_{1}}\eta^{i_{1}j_{1}}x_{j_{1}}= x_{i_{2}}\eta^{i_{2}j_{2}}x_{j_{2}}= -\frac{1}{2}r^{2}.\tag{2-18}
\end{equation}
which is the space $ AdS^{2}\times AdS^{2} $ with radius $ r/\sqrt{2} $. So $ X_{AdS}^{5} $ is a principal $ U(1) $ bundle over $ AdS^{2}\times AdS^{2} $:
\begin{equation}
U(1) \;\xrightarrow{right U(1)-action}\; X_{AdS}^{5}\;\xrightarrow{\pi}\; AdS^{2}\times AdS^{2}.\tag{2-19}
\end{equation}
Noncommutative geometry is a pointless geometry. In this geometry instead of the coordinates $ (x_{i_{1}}, x_{i_{2}} )$ of $ AdS^{2}\times AdS^{2} $, the $ SU(1,1)\times SU(1,1) $ angular momentum generators in the unitary  irreducible $ l_{1},l_{2}-$representation spaces have the role of the points of the fuzzy $ AdS^{2}\times AdS^{2} $ i.e. $ AdS_{F}^{2}\times AdS_{F}^{2} $. Let us consider $ X_{i_{1}}= \mu_{1} L_{i_{1}} $ and $ X_{i_{2}}= \mu_{2} L_{i_{2}} $(here we use $ L_{i_{1}} $ and $ L_{i_{2}} $ as the generators of the first and second $ SU(1,1) $, respectively).$ L_{i_{1,2}} $ satisfies the $ su(1,1) $ Lie algebra:
\begin{equation}
[L_{i_{1}}, L_{j_{1}}]= iC_{i_{1}j_{1}}^{\;\;\;k_{1}}L_{k_{1}}, \quad [L_{i_{2}}, L_{j_{2}}]= iC_{i_{2}j_{2}}^{\;\;\; k_{2}}L_{k_{2}}, \quad [L_{i_{1}}, L_{i_{2}}]=0.\tag{2-20}
\end{equation}
$ \mu_{1,2} $ determined by the value of $ su(1,1) $ Casimir operator 
\begin{equation}
\dfrac{1}{\mu_{1}^{2}•}= C_{su(1,1)}= -l_{1}(l_{1}-1), \quad \dfrac{1}{\mu_{2}^{2}•}= C_{su(1,1)}= -l_{2}(l_{2}-1). \tag{2-21}
\end{equation}
Then, the noncommutative coordinates are
\begin{equation}
X_{i_{1}}= \dfrac{L_{i_{1}}}{2\sqrt{l_{1}(1-l_{1})}•}, \quad X_{i_{2}}= \dfrac{L_{i_{2}}}{•2\sqrt{l_{2}(1-l_{2})}},\tag{2-22}
\end{equation}
which satisfy the $ su(1,1) $ Lie algebra
\begin{equation}
[X_{i_{1}}, X_{j_{1}}]= \dfrac{i}{2\sqrt{l_{1}(1-l_{1})}•}C_{i_{1}j_{1}}^{\;\;\;k_{1}}X_{k_{1}},\quad 
[X_{i_{2}}, X_{j_{2}}]= \dfrac{i}{2\sqrt{l_{2}(1-l_{2})}•}C_{i_{2}j_{2}}^{\;\;\;k_{2}}X_{k_{2}}, \quad
 [X_{i_{1}}, X_{i_{2}}]= 0. \tag{2-23}
\end{equation}
In the principal fibration  $X_{AdS}^{5}\;\xrightarrow{U(1)}\;AdS^{2}\times AdS^{2}$, the module of sections is $C( AdS^{2}\times AdS^{2})$-module $\Gamma^{\infty}(AdS^{2}\times AdS^{2},E^{(n)})$ in which $C( AdS^{2}\times AdS^{2})$ is the commutative algebra of functions on the manifold $AdS^{2}\times AdS^{2}$. In the fuzzy case, this algebra is a noncommutative algebra and therefore, left and right modules are not isomorphic. In this case to each angular momentum operator $\mathbf{L}_{1,2}$, we associate two linear operators $\mathbf{L}_{1,2}^{L}$ and $\mathbf{L}_{1,2}^{R}$ with the left and right actions on the fuzzy pseudo Hermitian matrix algebra $\mathcal{A}_{l_{1},l_{2}}= \lbrace\psi\in M_{(2l_{1}+1)(2l_{2}+1)}(\mathbb{C})\rbrace $:
\begin{equation}
L_{i_{\alpha}}^{L} \psi=L_{i_{\alpha}} \psi ,\quad L_{i_{\alpha}}^{R} \psi=\psi L_{i_{\alpha}} ,\quad\forall\psi \in \mathcal{A}_{l_{1},l_{2}}, \alpha= 1,2,
\tag{2-24}
\end{equation}
where the right action satisfies the $ su(1,1) $ algebra with minus sign $ - L_{i_{\alpha}}^{R} $
.These left and right operators commute with each other:
\begin{equation}
[L_{i_{\alpha}}^{L},L_{j_{\alpha}}^{R}]=0, \quad \alpha= 1,2.
\tag{2-25}
\end{equation}
The $ \mathbf{L}_{\alpha}^{L}$ and $ \mathbf{L}_{\alpha}^{R} $ have the same $ su(1,1) $ algebra.
The coordinates $ (x_{i_{1}}, x_{i_{2}}) $ of commutative $ AdS^{2}\times AdS^{2} $ can be obtain as the limit case
\begin{equation}
 x_{i_{\alpha}}=\lim_{l_{\alpha} \to \infty}\frac{L_{i_{\alpha}}^{L,R}}{\sqrt{l_{\alpha}(1-l_{\alpha})}}  =\lim_{l_{\alpha} \to \infty}\dfrac{L_{i_{\alpha}}^{L,R}}{l_{\alpha}}.
\tag{2-26}
\end{equation}
We use $ \mathbf{L}_{\alpha}^{L} $, $ \mathbf{L}_{\alpha}^{R} $, to define the fuzzy version of orbital momentum operators $\boldsymbol{\mathcal{L}_{\alpha}}$ on the fuzzy  $ AdS_{F}^{2}\times AdS_{F}^{2}$. We define $\boldsymbol{\mathcal{L}_{\alpha}}$ by the adjoint action of $ L_{i_{\alpha}} $ on the $\mathcal{A}_{l_{\alpha}}$:
\begin{equation}
\mathcal{L}_{i_{\alpha}} \psi =(L_{i_{\alpha}}^{L} - L_{i_{\alpha}}^{R})\psi= ad_{L_{i_{\alpha}}}\psi = [L_{i_{\alpha}},\psi], \quad \alpha= 1,2.
\tag{2-27}
\end{equation}
It is easy to see that the algebra of $ \mathcal{L}_{i_{\alpha}} $ is $ su(1,1) $ Lie algebra:
\begin{equation}
[\mathcal{L}_{i_{\alpha}}, \mathcal{L}_{j_{\alpha}}]= iC_{i_{\alpha}j_{\alpha}}^{\;\;\;k_{\alpha}}\mathcal{L}_{k_{\alpha}}, \quad [\mathcal{L}_{i_{1}}, \mathcal{L}_{i_{2}}]= 0.\tag{2-28}
\end{equation}
In the commutative limit we have the following angular momentum operators on commutative anti-deSitter space:
\begin{equation}
\lim_{l_{\alpha} \to \infty}( L_{i_{\alpha}}^{L} - L_{i_{\alpha}}^{R})= iC_{i_{\alpha}j_{\alpha}}^{\;\;\;k_{\alpha}} x_{j_{\alpha}}\dfrac{\partial}{•\partial x_{k_{\alpha}}}, \quad \alpha=1,2.\tag{2-29}
\end{equation}

\section{Projective module description of the fibration $ X_{AdS}^{5}\rightarrow AdS^{2}\times AdS^{2} $}
Consider the $U(1)$ principal fibration  $\pi$ with $X_{AdS}^{5} \cong \dfrac{SU(1,1)\times SU(1,1)}{U(1)•}$ as total space:
\begin{equation}
U(1) \;\xrightarrow{right U(1)-action}\; X_{AdS}^{5}\;\xrightarrow{\pi}\; AdS^{2}\times AdS^{2},
\tag{3-1}
\end{equation}
over the four-dimensional cross manifold $ AdS^{2}\times AdS^{2}$. The total manifold is $ X_{AdS}^{5} $ which is the base of the conifold $ Y_{AdS}^{6} $.
Let $B_{\mathbb{C}} =C^{\infty}(X_{AdS}^{5},\mathbb{C})$ and $A_{\mathbb{C}}= C^{\infty}(AdS^{2}\times AdS^{2},\mathbb{C})$ denote the algebras of $\mathbb{C}$-valued smooth functions on the total manifold $X_{AdS}^{5}$ and base manifold $ AdS^{2}\times AdS^{2}$ under point-wise multiplication, respectively. The irreducible representations of the group $ U(1) $ are labeled by an integer $ n $. The elements of $B_{\mathbb{C}} $ can be classified into the right modules,
\begin{equation}
C_{(\pm n)}^{\infty}(X_{AdS}^{5},\mathbb{C}) =\{\varphi_{(\pm n)} :X_{AdS}^{5} \rightarrow \mathbb{C},\quad \varphi_{(\pm n)} (p\cdot\omega)= \omega^{(\pm n)} \cdot \varphi(p)\:, \quad \forall p \in X_{AdS}^{5}\: ,\:\forall\omega\in U(1)\}
\tag{3-2},
\end{equation}
over the pull back of the $A_{\mathbb{C}}$. The left actions of the group $ U(1) $ on $ \mathbb{C} $ are labeled by an integer $ n $ which characterizes the bundle. The Serre-Swan theorem [25] states that for a compact smooth manifold $ M $, there is a complete  equivalence between the category of vector bundles over that manifold and bundle maps, and the category of finitely generated projective modules over the algebra $ C(M) $ of functions over $ M $ and module morphisms. In algebraic $ K $-theory, it is well known that corresponds to these bundles, there are  projectors $ P_{n} $ [25] such that, for the associated  vector bundle 
\begin{equation}
E^{(n)} = X_{AdS}^{5} \times_{U(1)} \mathbb{C} \xrightarrow{\pi} AdS^{2}\times AdS^{2},
\tag{3-3}
\end{equation}
right $ A_{\mathbb{C}} $-module of sections $\Gamma^{\infty}(AdS^{2}\times AdS^{2} ,E^{(n)})$ which is isomorphic with  $C_{(n)}^{\infty}(X_{AdS}^{5}, \mathbb{C})$ is equivalent to the image in the free module $ (A_{\mathbb{C}})^{(2n+1)}=C^{\infty}( AdS^{2}\times AdS^{2},\mathbb{C}) \otimes \mathbb{C}^{2n+1} $ of a projector $ P_{n} $, $\Gamma^{\infty}( AdS^{2}\times AdS^{2},E^{(n)})=P_{n}(A_{\mathbb{C}})^{2n+1} $. The projector $ P_{n} $ is a $ \Lambda- $pseudo Hermitian operator of rank $1$ over $\mathbb{C} $
\begin{equation}
P_{n}\in M_{2n+1} (A_{\mathbb{C}}),\quad P_{n}^{2}=P_{n},\quad P_{n}^{\dagger} = \Lambda P_{n}\Lambda^{-1},\quad Tr P_{n}=1.
\tag{3-4}
\end{equation} 
where $ Tr $ is trace and$ 1 $ is the constant function.
For the right $ A_{\mathbb{C}} $-module of sections  $ \Gamma^{\infty}(AdS^{2}\times AdS^{2},E^{(n)}) $ there exist $ n+1 $ projectors $ P_{1},P_{2},....,P_{n+1} $ having the same rank $ 1 $. Therefore, the free module $ (A_{\mathbb{C}})^{2n+1} $ can be written as a direct sum of the projective $ A_{\mathbb{C}}- $modules,
\begin{equation}
(A_{\mathbb{C}})^{2n+1}=\bigoplus_{\substack{i=1}}^{\substack{2n+1}} P_{i}(A_{\mathbb{C}})^{2n+1}.
\tag{3-5}
\end{equation}

\section{Spin $\frac{1}{2}$  pseudo-projectors of the pseudo-projective \\ $ A(AdS_{F}^{2}\times AdS_{F}^{2}) -$module}

According to the Serre-Swan's theorem, in noncommutative geometry, the study of the  principal fibration \\ $X_{AdS_{F}}^{5}\xrightarrow{U(1)} AdS_{F}^{2}\times AdS_{F}^{2}$, replaces with the study of noncommutative finitely generated pseudo-projective $ A(AdS_{F}^{2}\times AdS_{F}^{2})  -$module of its sections. To build the left and right pseudo-projective modules we should construct the fuzzy pseudo-projectors of these modules.
The pseudo $ \Lambda- $projectors for left pseudo-projective module can be written as:
\begin{equation}
P_{l_{\alpha}\pm\frac{1}{2}}^{L,\alpha} = \frac{1}{2}\left\lbrace 1\mp\frac{(\mathbf{\Sigma}_{\alpha}\cdot \mathbf{X}_{\alpha}-\frac{\mu_{\alpha}}{2})}{\sqrt{1-\frac{  \mu_{\alpha}^{2}}{4}}}\right\rbrace,\quad (P_{l_{\alpha}\pm\frac{1}{2}}^{L,\alpha})^{\dagger}= \Lambda P_{l_{\alpha}\pm\frac{1}{2}}^{L,\alpha}\Lambda^{-1}, \alpha=1,2, \tag{4-1}
\end{equation}
where $ \alpha=1 $ and $ \alpha=2 $ are for the first and second fuzzy anti-deSitter space in $ AdS_{F}^{2}\times AdS_{F}^{2} $, respectively and $ \Sigma_{\alpha,1}, \Sigma_{\alpha,2} $ and $ \Sigma_{\alpha,3} $ are the generators of $ su(1,1) $ Lie algebra.
substituting (2-21) in (4-1) we can write:
\begin{equation}
P_{(l_{\alpha}\pm 1/2)}^{L,\alpha}= \dfrac{1}{2•}[1\mp \dfrac{\mathbf{\Sigma}_{\alpha}\cdot \mathbf{L}_{\alpha}^{L}-1}{\sqrt{l_{\alpha}(1-l_{\alpha})-1}•}],
\tag{4-2}
\end{equation}
which couples left angular momentum and spin $ \dfrac{1}{2} $ to its maximum and minimum values $ \l_{\alpha}\pm \dfrac{1}{2} $, respectively.
It is easy to see that
\begin{equation}
P_{(l_{\alpha}+ \frac{1}{2})}^{L,\alpha} + P_{(l_{\alpha}-\frac{1}{2})}^{L,\alpha} = 1_{(2l_{\alpha}+1)(2l_{\alpha}+1)}. 
\tag{4-3}
\end{equation}
These are the pseudo-projectors of our left projective $ A(AdS_{F}^{2})- $module and we have
\begin{equation}
(A(AdS_{F}^{2}))^{2}=(A(AdS_{F}^{2}))^{2}P_{(l_{\alpha}+ \frac{1}{2})}^{L,\alpha}\oplus (A(AdS_{F}^{2}))^{2}P_{(l_{\alpha}- \frac{1}{2})}^{L,\alpha}.\tag{4-4}
\end{equation}
 One can expand these operators to act on $ A(AdS_{F}^{2}\times AdS_{F}^{2})  -$module. Let us define $ P^{L}:= (P^{L,1}, P^{L,2}) $. Then we have 
\begin{equation}
( A(AdS_{F}^{2}\times AdS_{F}^{2}) )^{2}=( A(AdS_{F}^{2}\times AdS_{F}^{2}) )^{2}P_{(l+ \frac{1}{2})}^{L}\oplus ( A(AdS_{F}^{2}\times AdS_{F}^{2}) )^{2}P_{(l- \frac{1}{2})}^{L}.\tag{4-5}
\end{equation}
Using (4-3) we can define the corresponding $ \Lambda- $pseudo idempotents as:
\begin{equation}
\Gamma_{(l_{\alpha}\pm \frac{1}{2})}^{L,\alpha}=2P_{(l_{\alpha}\pm \frac{1}{2})}^{L,\alpha}-1= \mp \frac{\mathbf{\Sigma}_{\alpha} \cdot \mathbf{L}_{\alpha}^{L}-1}{\sqrt{l_{\alpha}(1-l_{\alpha})-1}},\quad ,(\Gamma_{(l_{\alpha}\pm \frac{1}{2})}^{L,\alpha})^{\dagger}= \Lambda \Gamma_{(l_{\alpha}\pm \frac{1}{2})}^{L,\alpha}\Lambda^{-1}.
\tag{4-6}
\end{equation}
 The pseudo-projectors $P_{(l_{\alpha}\pm \frac{1}{2})}^{R,\alpha} $ coupling the right momentum and spin $ \dfrac{1}{2} $ to its maximum and minimum values $ l_{\alpha}\pm \dfrac{1}{2} $, respectively, are obtained by changing $ L_{i_{\alpha}}^{L} $ to $ -L_{i_{\alpha}}^{R} $ in the above expression
 \begin{equation}
P_{(l_{\alpha}\pm 1/2)}^{R,\alpha}= \dfrac{1}{2•}[1\pm \dfrac{\mathbf{\Sigma}_{\alpha}\cdot \mathbf{L}_{\alpha}^{R}+1}{\sqrt{l_{\alpha}(1-l_{\alpha})-1}•}],\quad  (P_{(l_{\alpha}\pm 1/2)}^{R,\alpha})^{\dagger}= \Lambda P_{(l_{\alpha}\pm 1/2)}^{R,\alpha}\Lambda^{-1}.
\tag{4-7}
\end{equation}
 These are the pseudo-projectors of our right projective $ A(AdS_{F}^{2})  -$module 
\begin{equation}
(A(AdS_{F}^{2}))^{2}=(A(AdS_{F}^{2}))^{2}P_{(l_{\alpha}+ \frac{1}{2})}^{R,\alpha}\oplus (A(AdS_{F}^{2}))^{2}P_{(l_{\alpha}- \frac{1}{2})}^{R,\alpha}.\tag{4-8}
\end{equation}
 One can expand these operators to act on $ A(AdS_{F}^{2}\times AdS_{F}^{2})  -$module. Let us define $ P^{R}:= (P^{R,1}, P^{R,2}) $. Then we have 
\begin{equation}
( A(AdS_{F}^{2}\times AdS_{F}^{2}) )^{2}=( A(AdS_{F}^{2}\times AdS_{F}^{2}) )^{2}P_{(l+ \frac{1}{2})}^{R}\oplus ( A(AdS_{F}^{2}\times AdS_{F}^{2}) )^{2}P_{(l- \frac{1}{2})}^{R}.\tag{4-9}
\end{equation}
 The corresponding $ \Lambda- $pseudo idempotents are
 \begin{equation}
\Gamma_{(l_{\alpha}\pm \frac{1}{2})}^{R,\alpha}=2P_{(l_{\alpha}\pm \frac{1}{2})}^{R,\alpha}-1= \pm \frac{\mathbf{\Sigma}_{\alpha} \cdot \mathbf{L}_{\alpha}^{R}+1}{\sqrt{l_{\alpha}(1-l_{\alpha})-1}}, \quad (\Gamma_{(l_{\alpha}\pm \frac{1}{2})}^{R,\alpha})^{\dagger}= \Lambda \Gamma_{(l_{\alpha}\pm \frac{1}{2})}^{R,\alpha}\Lambda^{-1} .
\tag{4-10}
\end{equation}

\section{Fuzzy pseudo Ginsparg-Wilson algebra and its Spin $\frac{1}{2}$ fuzzy Dirac and chirality operators on $ AdS_{F}^{2}\times AdS_{F}^{2} $}
 The fuzzy Ginsparg-Wilson algebra $ \mathcal{A}  $ is the $ \Lambda- $pseudo $ \dagger $ -algebra  over $ \mathbb{C} $, generated by two $ \Lambda- $pseudo $ \dagger $ -invariant involution $ \Gamma_{\alpha} $ and ${\Gamma_{\alpha}^{\prime}}$:
\begin{equation}
\mathcal{A}_{\alpha} = \langle\Gamma_{\alpha},{\Gamma_{\alpha}^{\prime}}\colon\quad(\Gamma_{\alpha})^{2} =({\Gamma_{\alpha}^{\prime}})^{2}=\textit{I}_{(2l_{\alpha}+1)(2l_{\alpha}+1)},\quad (\Gamma_{\alpha})^{\dagger}=\Lambda\Gamma_{\alpha}\Lambda^{-1},\quad ({\Gamma_{\alpha}^{\prime}})^{\dagger}=\Lambda{\Gamma_{\alpha}^{\prime}}\Lambda^{-1}\rangle , \alpha=1,2,
\tag{5-1}
\end{equation}
where $ \alpha=1,2 $ denote the Ginsparg-Wilson algebra associated to first and second anti-deSitter in $ AdS_{F}^{2}\times AdS_{F}^{2} $.
Each representation of (5-1) is a realization of the Ginsparg-Wilson algebra.
Now, consider the following two elements constructed out of the generators $ \Gamma_{\alpha} $ and ${\Gamma_{\alpha}^{\prime}}  $ of the pseudo fuzzy Ginsparg-Wilson algebra $ \mathcal{A_{\alpha}} $:
\begin{equation}
\begin{split}
\Gamma_{\alpha,1}= \frac{1}{2} (\Gamma_{\alpha} + {\Gamma_{\alpha}^{\prime}}) \; , \qquad \qquad {(\Gamma_{\alpha,1}})^{\dagger} =\Lambda \Gamma_{\alpha,1}\Lambda^{-1} ,\\
\Gamma_{\alpha,2} = \frac{1}{2} (\Gamma_{\alpha} - {\Gamma_{\alpha}^{\prime}}) \; , \qquad \qquad {(\Gamma_{\alpha,2}})^{\dagger} = \Lambda \Gamma_{\alpha,2}\Lambda^{-1} .
\end{split}
\tag{5-2}
\end{equation}
So that, $\Gamma_{\alpha,1}$ and $\Gamma_{\alpha,2}$ anticommute with each other:
\begin{equation}
\left\lbrace \Gamma_{\alpha,1}, \Gamma_{\alpha,2}\right\rbrace  = 0.
\tag{5-3}
\end{equation}
Identifying $ \Gamma_{(l_{\alpha}\pm \frac{1}{2})} ^{L,\alpha}$ and $ \Gamma_{(l_{\alpha}\pm \frac{1}{2})} ^{R,\alpha}$ with $ \Gamma_{\alpha} $ and ${\Gamma_{\alpha}^{\prime}} $, we get:
\begin{equation}
\Gamma_{\alpha,1}^{\pm}= \pm \frac{\mathbf{\Sigma}_{\alpha} \cdot \mathbf{\mathcal{L}}_{\alpha}^{F}-1}{\sqrt{l_{\alpha}(1-l_{\alpha})-1}}, \qquad 
\Gamma_{\alpha,2}^{\pm }= \pm \frac{\mathbf{\Sigma}_{\alpha} \cdot (\mathbf{L}_{\alpha}^{L}+\mathbf{L}_{\alpha}^{R})}{\sqrt{l_{\alpha}(1-l_{\alpha})-1}}.
\tag{5-4}
\end{equation}
Now, let us define the $ \Lambda- $pseudo fuzzy Dirac operator on each fuzzy anti-deSitter spaces $ AdS_{F}^{2}$ as:
\begin{equation}
D_{\alpha,F}^{\pm }= \sqrt{l_{\alpha}(1-l_{\alpha})-1} \Gamma_{\alpha,1}^{\pm }= \pm (\mathbf{\Sigma}_{\alpha} \cdot \mathcal{L}_{\alpha}^{F}-1)= \pm(\Sigma_{i_{\alpha}}\eta^{i_{\alpha}j_{\alpha}}\mathcal{L}_{j_{\alpha}}^{F}), \quad (D_{\alpha,F}^{\pm })^{\dagger}= \Lambda D_{\alpha,F}^{\pm }\Lambda^{-1}.
\tag{5-5}
\end{equation}
In the limit case (5-5)becomes the Dirac operator on each commutative $ AdS^{2} $:
\begin{equation}
 \lim_{l_{\alpha} \to \infty } D_{\alpha,F}^{\pm }= \pm (\mathbf{\Sigma}_{\alpha} \cdot \mathcal{L}_{\alpha} - 1).
 \tag{5-6}
\end{equation}
Also, we can define pseudo-chirality operators on $ AdS_{\alpha,F}^{2} $ as $ \gamma_{\alpha,F}^{\pm }= \Gamma_{\alpha,2}^{\pm } $ which in the commutative limit they become:
\begin{equation}
 \lim_{l_{\alpha}\to \infty} \gamma_{\alpha,F}^{\pm }=\pm \mathbf{\Sigma}_{\alpha}\cdot \mathbf{x}_{\alpha}= \pm \Sigma_{i_{\alpha}}\eta^{i_{\alpha}j_{\alpha}}x_{j_{\alpha}}.
 \tag{5-7}
\end{equation}
Also, it is easy to see that 
\begin{equation}
 \lim_{l_{\alpha}\to \infty} \lbrace D_{\alpha,F}^{\pm },\gamma_{\alpha,F}^{\pm }\rbrace=0
 \tag{5-8}
\end{equation}
which we expect from Dirac and chirality operators on each $ AdS^{2} $.
Now, let us define the Ginsparg-Wilson algebra for $ AdS_{F}^{2}\times AdS_{F}^{2} $ and then construct its corresponding pseudo Dirac and chirality operators. This algebra is a $ \Lambda- $pseudo$ \dagger- $invariant algebra over $ \mathbb{C} $ which can be defined as:
\begin{equation}
\mathcal{A}_{l_{1},l_{2}}= \langle \Gamma= \Gamma_{1}\Gamma_{2},\quad \Gamma^{'}= \Gamma_{1}^{'}\Gamma_{2}^{'} \quad \Gamma^{2}= \Gamma^{'2}= 1, \quad \Gamma^{\dagger}= \Lambda\Gamma \Lambda^{-1}, \quad \Gamma^{'\dagger}=\Lambda \Gamma^{'}\Lambda^{-1} \rangle.\tag{5-9}
\end{equation}
where $ \Gamma_{1,2} $ and $ \Gamma_{1,2}^{'} $ are the generators of the Ginsparg-Wilson algebra of each $ AdS_{F}^{2} $ which are given in (5-4). Also, they satisfy the following commutation relations
\begin{equation}
[\Gamma_{1}, \Gamma_{2}]= [\Gamma_{1}, \Gamma_{2}^{'}]= [\Gamma_{1}^{'}, \Gamma_{2}]=0.\tag{5-10}
\end{equation}
We consider the radii of the anti-deSitter spaces equal. We can chose one of the generators of the Ginsparg-Wilson algebra (5-9) as chirality operator on $ AdS_{F}^{2}\times AdS_{F}^{2} $ because in the commutative limit both of them become the same chirality operators $ \gamma= \gamma_{1}\gamma_{2} $ on the commutative $ AdS^{2}\times AdS^{2} $. Let us define the $ \Lambda- $pseudo fuzzy Dirac operator as:
\begin{equation}
D_{F}= \dfrac{\Gamma_{1}\Gamma_{2}-\Gamma_{1}^{'}\Gamma_{2}^{'}}{2\sqrt{l_{1}(1-l_{1})-1}\sqrt{l_{2}(1-l_{2})-1}•}, \quad D_{F}^{\dagger}=\Lambda D_{F}\Lambda^{-1}\tag{5-11}
\end{equation}
It is easy to see that
\begin{equation}
\Gamma_{1}\Gamma_{2}-\Gamma_{1}^{'}\Gamma_{2}^{'}= \dfrac{1}{2•}[(\Gamma_{1}- \Gamma_{1}^{'})(\Gamma_{2} +\Gamma_{2}^{'})+ (\Gamma_{1}+\Gamma_{1}^{'})(\Gamma_{2}- \Gamma_{2}^{'})]\tag{5-12}
\end{equation}
Now, using the following definitions
\begin{equation}
D_{1}^{F}= \dfrac{ (\Gamma_{1}- \Gamma_{1}^{'})(\Gamma_{2} +\Gamma_{2}^{'}) }{2•\sqrt{l_{1}(1-l_{1})-1}\sqrt{l_{2}(1-l_{2})-1}}, \quad D_{2}^{F}= \dfrac{ (\Gamma_{1}+ \Gamma_{1}^{'})(\Gamma_{2} -\Gamma_{2}^{'}) }{2•\sqrt{l_{1}(1-l_{1})-1}\sqrt{l_{2}(1-l_{2})-1}}\tag{5-13}
\end{equation}
which satisfies
\begin{equation}
[D_{1}^{F}, D_{2}^{F}]=0, \quad (D_{1,2}^{F})^{\dagger}= \Lambda D_{1,2}^{F}\Lambda^{-1}.\tag{5-14}
\end{equation}
The pseudo fuzzy Dirac operator (5-11) on $ AdS_{F}^{2}\times AdS_{F}^{2} $ can be written as:
\begin{equation}
D_{F}= D_{1}^{F}+D_{2}^{F}.\tag{5-15}
\end{equation}
In the commutative limit (5-15) tends to 
\begin{equation}
\lim_{l_{1,2} \to \infty }D_{F}= D_{1}+D_{2}= (\mathbf{\Sigma}_{1}\cdot \mathcal{L}_{1}+1)(\mathbf{\Sigma}_{2}\cdot \mathbf{x}_{i_{2}})+(\mathbf{\Sigma}_{1}\cdot \mathbf{x}_{i_{1}})(\mathbf{\Sigma}_{2}\cdot \mathcal{L}_{2}+1).\tag{5-16}
\end{equation}

\section{ Fuzzy gauged pseudo Dirac operator ( no instanton fields)}
Let us denote by $ A^{L}=(A_{i_{1}}^{L}, A_{i_{2}}^{L}) $ the  connection $ 1- $form associated with the pseudo-projector $ P $ on $ AdS_{F}^{2}\times AdS_{F}^{2} $. $ A_{i_{1}}^{L} $ and $ A_{i_{2}}^{L} $ are connection $ 1- $forms on the first and second $ AdS_{F}^{2} $, respectively.
\begin{equation}
 A^{L}\in End_{\mathbb{C}}( C^{\infty}(X_{AdS_{F}}^{5}),\mathbb{C})\otimes_{\mathbb{C}}\Omega^{1}(( X_{AdS_{F}}^{5}),\mathbb{C})
 \tag{6-1}
\end{equation}
The components of this $ U(1) $ gauged field according to our principal fibration are given by 
\begin{equation}
 A= (dx_{i_{1}}A_{i_{1}}, dx_{i_{2}}A_{i_{2}}).
 \tag{6-2}
\end{equation}
The $ \dagger- $invariant fuzzy gauge field $ A^{L} $ acts on $ \xi = ( \xi_{i_{1}} ,\xi_{i_{2}}),\xi_{i} \in  AdS^{2}_{F}  (2l + 1) $ as:
\begin{equation}
[(A_{i_{1}}^{L}, A_{i_{2}}^{L})(\xi_{i_{1}},\xi_{i_{2}} )]_{m}=((A_{i_{1}})_{mn}(\xi_{i_{1}})_{n}, (A_{i_{2}})_{mn}(\xi_{i_{2}})_{n}).
\tag{6-3}
\end{equation}
The $ \Lambda- $pseudo $ \dagger $-invariant condition on $ A_{i_{1},i_{2}}^{L} $ is:
\begin{equation}
(A_{i_{1},i_{2}}^{L})^{\dagger} =\Lambda A_{i_{1},i_{2}}^{L}\Lambda^{-1}.
\tag{6-4}
\end{equation}
The corresponding curvature 2-form $ F^{L} $ on $ X_{AdS_{F}}^{5} $ 
 \begin{equation}
 F^{L}\in End_{\mathbb{C}}( C^{\infty}(X_{AdS_{F}}^{5}),\mathbb{C})\otimes_{\mathbb{C}}\Omega^{2}(( X_{AdS_{F}}^{5}),\mathbb{C})
 \tag{6-5}
\end{equation}
is given by
\begin{equation*}
 F_{i_{1}j_{1}}^{L}= i([L_{i_{1}}^{L}, A_{j_{1}}^{L}]+ [A_{i_{1}}^{L}, L_{j_{1}}^{L}]+ [A_{i_{1}}^{L}, A_{j_{1}}^{L}]), 
\end{equation*}
\begin{equation*}
 F_{i_{2}j_{2}}^{L}= i([L_{i_{2}}^{L}, A_{j_{2}}^{L}]+ [A_{i_{2}}^{L}, L_{j_{2}}^{L}]+ [A_{i_{2}}^{L}, A_{j_{2}}^{L}]), 
\end{equation*}
\begin{equation}
 F_{i_{1}j_{2}}^{L}= i([L_{i_{1}}^{L}, A_{j_{2}}^{L}]+ [A_{i_{1}}^{L}, L_{j_{2}}^{L}]+ [A_{i_{1}}^{L}, A_{j_{2}}^{L}]). 
 \tag{6-6}
\end{equation}
The components of curvature 2-form also, satisfies:
\begin{equation*}
[(L+ A)_{i_{1}}^{L}, (L+ A)_{j_{1}}^{L}]= iC_{i_{1}j_{1}}^{\;\;\;k_{1}}(L+ A)_{k_{1}}^{L}+ iF_{i_{1}j_{1}}^{L},
\end{equation*}
\begin{equation*}
 [(L+ A)_{i_{1}}^{L}, (L+ A)_{j_{2}}^{L}]= iC_{i_{1}j_{2}}^{\;\;\;k_{\alpha}}(L+ A)_{k_{\alpha}}^{L}+ iF_{i_{1}j_{2}}^{L},
\end{equation*}
\begin{equation}
[(L+ A)_{i_{2}}^{L}, (L+ A)_{j_{2}}^{L}]= iC_{i_{2}j_{2}}^{\;\;\;k_{\alpha}}(L+ A)_{k_{\alpha}}^{L}+ iF_{i_{2}j_{2}}^{L},
 \tag{6-7}
\end{equation}

The fuzzy gauge field $ A^{L} $ on the commutative $ AdS^{2}\times AdS^{2} $ becomes a commutative field $ \mathbf{a}=(\mathbf{a}_{1}, \mathbf{a}_{2})$ and its components $ a_{i_{1,2}} $satisfies the following condition: 
\begin{equation}
\mathbf{x}\cdot \mathbf{a}=(x_{i_{1}}\eta^{i_{1}j_{1}}a_{j_{1}}, x_{i_{2}}\eta^{i_{2}j_{2}}a_{j_{2}})= (0, 0).
\tag{6-8}
\end{equation}
We need a condition to get the above result for large $ l $. One of the conditions of such a nature on each $ AdS^{2} $ is :
\begin{equation}
(\mathbf{L}_{\alpha}^{L} + \mathbf{A}_{\alpha}^{L}) \cdot (\mathbf{L}_{\alpha}^{L} + \mathbf{A}_{\alpha}^{L})= \mathbf{L}_{\alpha}^{L} \cdot \mathbf{L}_{\alpha}^{L} = L_{i_{\alpha}}^{L}\eta^{i_{\alpha}j_{\alpha}}L_{j_{\alpha}}^{L}= l_{\alpha}(1-l_{\alpha}), \quad \alpha=1,2.
\tag{6-9}
\end{equation}
The expansion of (6-9) is:
\begin{equation}
 L_{i_{\alpha}}^{L}\eta^{i_{\alpha}j_{\alpha}}A_{j_{\alpha}}^{L}+ A_{i_{\alpha}}^{L}\eta^{i_{\alpha}j_{\alpha}}L_{j_{\alpha}}^{L}+A_{i_{\alpha}}^{L}\eta^{i_{\alpha}j_{\alpha}} A_{j_{\alpha}}^{L}=0.
\tag{6-10}
\end{equation}
When the parameter $ l_{\alpha} $ tends to infinity, $ \dfrac{A_{i_{\alpha}}^{L}}{l_{\alpha}} $ tends to zero.
 Also, in this limit $ L_{i_{1}}^{L} $, $ L_{i_{2}}^{L} $ and $ A_{i,\alpha}^{L} $ tends to $ x_{i_{1}}$, $ x_{i_{2}} $and $ a_{i_{\alpha}} $, respectively . So we have the condition $ \mathbf{x}_{\alpha} \cdot \mathbf{a}_{\alpha}=0 $ on each $ AdS^{2} $.\\
Now, we can introduced the $ \Lambda- $pseudo gauged Ginsparg-Wilson system on each $ AdS^{2} $ as follow: For each $ \alpha=1,2 $, we can set:
\begin{equation}
\mathcal{A}_{\alpha}^{\pm}(\mathbf{A}_{\alpha}^{L})= \langle \Gamma_{\alpha}^{\pm}(\mathbf{A^{L}}), \Gamma_{\alpha}^{'} : \Gamma_{\alpha}^{\pm^{^{2}}}(\mathbf{A}^{L})=\Gamma_{\alpha}^{'^{2}}=1,\quad\Gamma_{\alpha}^{\pm \dagger}(\mathbf{A}^{L})=\Lambda\Gamma_{\alpha}^{\pm}(\mathbf{A}^{L})\Lambda^{-1} ,\quad\Gamma_{\alpha}^{'\dagger}=\Lambda\Gamma_{\alpha}^{'}\Lambda^{-1} \rangle, 
\tag{6-11}
\end{equation}
where
\begin{equation}
\Gamma_{\alpha}^{\pm }(\mathbf{A}^{L}) = \mp \dfrac{\mathbf{\Sigma}_{\alpha}\cdot (\mathbf{L}_{\alpha}^{L} + \mathbf{A}_{\alpha}^{L})-1}{|\mathbf{\Sigma}_{\alpha}\cdot (\mathbf{L}_{\alpha}^{L} + \mathbf{A}_{\alpha}^{L})-1|},\quad \Gamma_{\alpha}^{' \pm }(\mathbf{A}_{\alpha}^{L}) = \Gamma_{\alpha}^{' \pm }(0)=\pm \dfrac{\mathbf{\Sigma}_{\alpha}\cdot \mathbf{L}_{\alpha}^{R}+1}{|\mathbf{\Sigma}_{\alpha}\cdot \mathbf{L}_{\alpha}^{L}+1|}
\tag{6-12}
\end{equation}
They are involutory and $ \Lambda- $pseudo $ \dagger $-invariant operators:
\begin{equation}
\Gamma_{\alpha}(\mathbf{A}_{\alpha}^{L})^{2} = 1 ,\qquad \Gamma_{\alpha}(\mathbf{A}_{\alpha}^{L})^{\dagger} = \Lambda\Gamma_{\alpha}(\mathbf{A}_{\alpha}^{L})\Lambda^{-1}.
\tag{6-13}
\end{equation}
The gauged involution (6-12), reduces to (5-4) for zero $ \mathbf{A}^{L}$. We put $ \Gamma_{\alpha} = \Gamma_{\alpha}(\mathbf{A}_{\alpha}^{L}=0) $.\\ Also, we can define the second gauged involution as:
\begin{equation}
\Gamma_{\alpha}^{'}(\mathbf{A}_{\alpha}^{L}) = \Gamma_{\alpha}^{'}(0)=\Gamma_{\alpha}^{'}.
\tag{6-14}
\end{equation}
We put $ \Gamma_{\alpha}^{'}=\Gamma_{\alpha}^{'}(A_{\alpha}^{L}=0) $. Notice that, the operators $ \mathbf{L}_{\alpha}^{L,R} $  do not have continuum limit as their squares $ l_{\alpha}(1-l_{\alpha})$ diverge as $ l_{\alpha} $ tends to infinity. In contrast, $\boldsymbol{\mathcal{L}}_{\alpha} $ and $ \mathbf{A}_{\alpha}^{L} $ do have continuum limits.\\
It is easy to see that up to the first order (6-12) becomes:
\begin{equation}
\Gamma_{\alpha}^{\pm} (\mathbf{A}_{\alpha}^{L}) = \mp \dfrac{\mathbf{\Sigma}_{\alpha}\cdot (\mathbf{L}_{\alpha}^{L} + \mathbf{A}_{\alpha}^{L})-1}{\sqrt{l_{\alpha}(1-l_{\alpha})-1}}.
\tag{6-15}
\end{equation}
and
\begin{equation}
\Gamma_{\alpha}^{' \pm } = \pm\dfrac{\mathbf{\Sigma}_{\alpha}\cdot \mathbf{L}_{\alpha}^{R} +1 }{\sqrt{l_{\alpha}(1-l_{\alpha})-1}}.
\tag{6-16}
\end{equation}
Using (6-15)and (6-16) we can construct the following $ \Lambda- $pseudo $ \dagger- $invariant operators:
\begin{equation}
\begin{split}
\Gamma_{1,\alpha}^{\pm }(A_{\alpha}^{L}) = \frac{1}{2} (\Gamma_{\alpha}^{\pm }(A_{\alpha}^{L}) + \Gamma_{\alpha}^{\prime \pm}) \; , \qquad \qquad (\Gamma_{1,\alpha}^{\pm })^{\dagger} = \Lambda\Gamma_{1,\alpha}^{\pm }\Lambda^{-1} ,\\
\Gamma_{2,\alpha}^{\pm }(A_{\alpha}^{L}) = \frac{1}{2} (\Gamma_{\alpha}^{\pm }(A_{\alpha}^{L}) - \Gamma_{\alpha}^{\prime \pm }) \; , \qquad \qquad (\Gamma_{2,\alpha}^{\pm })^{\dagger} =  \Lambda\Gamma_{2,\alpha}^{\pm }\Lambda^{-1}.
\end{split}
\tag{6-17}
\end{equation}
Now, let us define the gauged pseudo fuzzy Dirac and chirality operators on each $ AdS_{F}^{2} $ as:
\begin{equation}
D_{F,\alpha}^{\pm }(\mathbf{A}_{\alpha}^{L})=\sqrt{l_{\alpha}(1-l_{\alpha})-1}\Gamma_{1,\alpha}^{\pm }(\mathbf{A}_{\alpha}^{L})=\pm (\mathbf{\Sigma}_{\alpha} \cdot (\mathcal{L}_{F,\alpha}+\mathbf{A}_{\alpha}^{L})-1),\qquad \alpha=1,2
\tag{6-18}
\end{equation}
and for the chirality operator:
\begin{equation}
\gamma_{F,\alpha}^{\pm }(\mathbf{A}_{\alpha}^{L})=\Gamma_{2,\alpha}^{\pm }(\mathbf{A}_{\alpha}^{L})= \mp \dfrac{\mathbf{\Sigma}_{\alpha}\cdot (\mathbf{L}_{\alpha}^{L}+\mathbf{L}_{\alpha}^{R} + \mathbf{A}_{\alpha}^{L})}{\sqrt{l_{\alpha}(1-l_{\alpha})-1}}.
\tag{6-19}
\end{equation}
In the commutative limit when $ l_{\alpha} $ tends to infinity (6-18) and (6-19) become:
\begin{equation}
 \lim_{l_{\alpha} \to \infty} D_{F,\alpha}^{\pm }(A_{\alpha}^{L})=\pm (\mathbf{\Sigma}_{\alpha} \cdot (\mathcal{L}_{\alpha}+A_{\alpha}^{L})-1),\qquad  \lim_{l_{\alpha} \to \infty}\gamma_{F,\alpha}^{\pm }(A_{\alpha}^{L}) =\mp \mathbf{\Sigma}_{\alpha} \cdot \mathbf{x}_{\alpha}.
 \tag{6-20}
\end{equation}
These are the correct pseudo gauged Dirac and chirality operators on each commutative $ AdS^{2} $.

Now, let us construct fuzzy Ginsparg-Wilson algebra for $ AdS_{F}^{2}\times AdS_{F}^{2} $. It is a $ \Lambda- $pseudo $ \dagger- $invariant algebra over $ \mathbb{C} $,
\begin{equation}
\mathcal{A}_{l_{1},l_{2}}(\mathbf{A}^{L})= \langle \Gamma(\mathbf{A}^{L})= \Gamma_{1}(\mathbf{A}_{1}^{L})\Gamma_{2}(\mathbf{A}_{2}^{L}),\quad \Gamma^{'}= \Gamma_{1}^{'}\Gamma_{2}^{'}, \quad \Gamma^{2}= \Gamma^{'2}= 1, \quad \Gamma^{\dagger}= \Lambda\Gamma\Lambda^{-1}, \quad \Gamma^{'\dagger}=\Lambda \Gamma^{'}\Lambda^{-1} \rangle,\tag{6-21}
\end{equation}
where $ \Gamma_{1,2} $ and $ \Gamma_{1,2}^{'} $ are the generators of the Ginsparg-Wilson algebra of each $ AdS_{F}^{2} $ which are given in (6-15) and (6-16).
Let us define the fuzzy $ \Lambda- $pseudo gauged Dirac operator on $ AdS_{F}^{2}\times AdS_{F}^{2} $ as:
\begin{equation}
D_{F}(\mathbf{A}^{L})= \dfrac{\Gamma_{1}(\mathbf{A}_{1}^{L})\Gamma_{2}(\mathbf{A}_{2}^{L})-\Gamma_{1}^{'}\Gamma_{2}^{'}}{2\sqrt{l_{1}(1-l_{1})-1}\sqrt{l_{2}(1-l_{2})-1}•}, \quad D_{F}(\mathbf{A}^{L})^{\dagger}=\Lambda D_{F}(\mathbf{A}^{L})\Lambda^{-1}. \tag{6-22}
\end{equation}
It is easy to see that
\begin{equation}
\Gamma_{1}(\mathbf{A}_{1}^{L})\Gamma_{2}(\mathbf{A}_{2}^{L})-\Gamma_{1}^{'}\Gamma_{2}^{'}= \dfrac{1}{2•}[(\Gamma_{1}(\mathbf{A}_{1}^{L})- \Gamma_{1}^{'})(\Gamma_{2}(\mathbf{A}_{2}^{L}) +\Gamma_{2}^{'})+ (\Gamma_{1}(\mathbf{A}_{1}^{L})+\Gamma_{1}^{'})(\Gamma_{2}(\mathbf{A}_{2}^{L})- \Gamma_{2}^{'})]\tag{6-23}
\end{equation}
Now, using the following definitions
\begin{equation}
D_{1}^{F}(\mathbf{A}_{1}^{L})= \dfrac{ (\Gamma_{1}(\mathbf{A}_{1}^{L})- \Gamma_{1}^{'})(\Gamma_{2}(\mathbf{A}_{2}^{L}) +\Gamma_{2}^{'}) }{2•\sqrt{l_{1}(1-l_{1})-1}\sqrt{l_{2}(1-l_{2})-1}}, \quad D_{2}^{F}(\mathbf{A}_{2}^{L})= \dfrac{ (\Gamma_{1}(\mathbf{A}_{1}^{L})+ \Gamma_{1}^{'})(\Gamma_{2}(\mathbf{A}_{2}^{L}) -\Gamma_{2}^{'}) }{2•\sqrt{l_{1}(1-l_{1})-1}\sqrt{l_{2}(1-l_{2})-1}},\tag{6-24}
\end{equation}
which satisfies
\begin{equation}
[D_{1}^{F}(\mathbf{A}_{1}^{L}), D_{2}^{F}(\mathbf{A}_{2}^{L})]=0,\tag{6-25}
\end{equation}
the fuzzy Dirac operator (6-22) on $ AdS_{F}^{2}\times AdS_{F}^{2} $ can be written as:
\begin{equation}
D_{F}(\mathbf{A}^{L})= D_{1}^{F}(\mathbf{A}_{1}^{L})+D_{2}^{F}(\mathbf{A}_{2}^{L}).\tag{6-26}
\end{equation}
In the commutative limit (6-26) tends to 
\begin{equation}
\lim_{l_{1,2} \to \infty }D_{F}(\mathbf{A}^{L})= D_{1}(\mathbf{A}_{1})+D_{2}(\mathbf{A}_{2})= (\mathbf{\Sigma}_{1}\cdot (\mathcal{L}_{1}+ \mathbf{A}_{1})-1)(\mathbf{\Sigma}_{2}\cdot \mathbf{x}_{i_{2}})+(\mathbf{\Sigma}_{1}\cdot \mathbf{x}_{i_{1}})(\mathbf{\Sigma}_{2}\cdot (\mathcal{L}_{2}+\mathbf{A}_{2})-1).\tag{6-27}
\end{equation}

\section{Fuzzy pseudo Dirac and chirality operators on $AdS_{F}^{2}\times AdS_{F}^{2}$ in  instanton sector}
As we mentioned in section $2$, according to the Serre-Swan's theorem, in noncommutative geometry, the study of the principal fibration $X_{AdS_{F}}^{5}\xrightarrow{U(1)} AdS_{F}^{2}\times AdS_{F}^{2}$, replaces with the study of noncommutative finitely generated projective $ A(AdS_{F}^{2}\times AdS_{F}^{2})  -$module of its sections. To build the pseudo-projective module, let introduce  $\mathbb{C}^{2t_{1}+1}\times \mathbb{C}^{2t_{2}+1}$ carrying the $ t_{1},t_{2} $-representations of angular momentum of $ su(1,1)\times su(1,1)$.
Here, the algebra $ su(1,1)\times su(1,1) $ is generated by elements $ T_{i_{1}} $ and $ T_{i_{2}} $ satisfying the following relations:
\begin{equation}
[T_{i_{1}}, T_{j_{1}}]= iC_{i_{1}j_{1}}^{\;\;\;k_{1}}T_{k_{1}}, \quad [T_{i_{2}}, T_{j_{2}}]= iC_{i_{2}j_{2}}^{\;\;\;k_{2}}T_{k_{2}}, \quad [T_{i_{1}}, T_{i_{2}}]=0.\tag{7-1}
\end{equation}
Also, let $ P_{F}^{(l_{\alpha}+t_{\alpha})} $ be the pseudo-projector coupling left angular momentum operator $ \mathbf{L}_{\alpha}^{L}$ with $ \mathbf{T}_{\alpha} $ to produce maximum angular momentum $ l_{\alpha}+t_{\alpha} $ on each $ AdS_{F}^{2} $. We know that the image of a  projector on a free module is a projective module. Then, as $ Mat(2l_{\alpha}+1)^{2t_{\alpha}+1}=Mat (2l_{\alpha}+1)\otimes \mathbb{C}^{2t_{\alpha}+1} $ is a free module, therefore, $  P^{(l_{\alpha}+t_{\alpha})} Mat(2l_{\alpha}+1)^{2t_{\alpha}+1} $ is the fuzzy version of $ U(1) $ bundle on each $AdS_{F}^{2}$. Also, we can use the pseudo-projector $ P_{F}^{(l_{\alpha}-t_{\alpha})} $ to produce the projective module $ P_{_{F}}^{(l_{\alpha}-t_{\alpha})} Mat (2l_{\alpha}+1)^{2t_{\alpha}+1}$ to introduce the least angular momentum $ (l_{\alpha}-t_{\alpha}) $.\\ The $ \Lambda- $pseudo fuzzy projectors $P_{_{F}}^{l_{\alpha}\pm t_{\alpha}}$ corresponding to $ (l_{\alpha}\pm t_{\alpha})$-representations of $ su(1,1) $ can be written as:
\begin{equation}
P_{F}^{(l_{\alpha} \pm t_{\alpha})} = \dfrac{1}{2•}(1 \mp \dfrac{•\mathbf{\Sigma}_{\alpha}\cdot(\mathbf{L}_{\alpha}^{L}+\mathbf{T}_{\alpha})-1}{\sqrt{(l_{\alpha}\pm t_{\alpha})(1-l_{\alpha}\mp t_{\alpha})-1}•}),\quad P_{_{F}}^{(l_{\alpha} \pm t_{\alpha})^{\dagger}}=\Lambda P_{F}^{(l_{\alpha} \pm t_{\alpha})}\Lambda^{-1},
\tag{7-2}
\end{equation}
\begin{equation}
Mat(2l_{\alpha}+1)\otimes\mathbb{C}^{2t_{\alpha}+1}=(Mat(2l_{\alpha}+1)\otimes\mathbb{C}^{2t_{\alpha}+1})P_{F}^{(l_{\alpha}+t_{\alpha})}\oplus  (Mat(2l_{\alpha}+1)\otimes \mathbb{C}^{2t_{\alpha}+1})P_{F}^{(l_{\alpha}-t_{\alpha})}.
\tag{7-3}
\end{equation}
To set the fuzzy Ginsparg-Wilson system in instanton sector to each $ AdS_{F}^{2} $, we choose the following $ \Lambda- $pseudo $ \dagger-$invariant involution $ \Gamma_{\alpha} $ for the highest and lowest weights $ l_{\alpha}\pm t_{\alpha} $:
\begin{equation}
\Gamma_{\alpha}^{\pm}(\mathbf{T}_{\alpha})=2 P_{F}^{(l_{\alpha}\pm t_{\alpha})}-1=\mp \dfrac{\mathbf{\Sigma}_{\alpha}\cdot(\mathbf{L}_{\alpha}^{L}+\mathbf{T}_{\alpha})-1}{\sqrt{(l_{\alpha}\pm t_{\alpha})(1-l_{\alpha}\mp t_{\alpha})-1}•}),\quad (\Gamma_{\alpha}^{\pm}(\mathbf{T}_{\alpha}))^{2}=1,\quad \Gamma_{\alpha}^{\pm^{\dagger}}(\mathbf{T}_{\alpha})=\Lambda\Gamma_{\alpha}^{\pm}(\mathbf{T}_{\alpha})\Lambda^{-1}.
\tag{7-4}
\end{equation}
We choose $ \Gamma_{\alpha}^{'} $ as in (6-16).It is clear that $ \Gamma_{\alpha}^{\pm}(T=0)=\Gamma_{\alpha} $. On the module $( Mat(2l_{\alpha}+1)^{2t_{\alpha}+1}\otimes\mathbb{C}^{2})P^{(l_{\alpha} \pm t_{\alpha})} $ we have:
\begin{equation}
(\mathbf{L}_{\alpha}^{L}+\mathbf{T}_{\alpha})^{2} =(l_{\alpha}\pm t_{\alpha})(1-l_{\alpha}\mp t_{\alpha}).
\tag{7-5}
\end{equation}
Now, we can introduce our pseudo fuzzy Ginsparg-Wilson system in instanton sector for each $ AdS_{F}^{2} $ as:
\begin{equation}
\mathcal{A}_{\alpha}^{\pm}(\mathbf{T}_{\alpha})=\langle\; \Gamma_{\alpha}^{\pm}(\mathbf{T}_{\alpha}), \Gamma_{\alpha}^{'} : \Gamma_{\alpha}^{\pm^{^{2}}}(\mathbf{T}_{\alpha})=\Gamma_{\alpha}^{'^{2}}=1,\quad\Gamma_{\alpha}^{\pm \dagger}(\mathbf{T}_{\alpha})=\Lambda\Gamma_{\alpha}^{\pm}(\mathbf{T}_{\alpha})\Lambda^{-1} ,\quad\Gamma_{\alpha}^{'\dagger}=\Lambda\Gamma_{\alpha}^{'}\Lambda^{-1} \rangle.
\tag{7-6}
\end{equation}
Using the definitions (5-2), (5-5) and (7-4) one can calculate $ \Lambda- $pseudo fuzzy Dirac and chirality operators on each $ AdS_{F}^{2} $ which in the commutative limit they become:
\begin{equation}
 \lim_{l_{\alpha} \to \infty} D_{F,\alpha}^{\pm }(\mathbf{T}_{\alpha})=\pm (\mathbf{\Sigma}_{\alpha} \cdot (\mathcal{L}_{\alpha}+\mathbf{T}_{\alpha})-1),\qquad  \lim_{l_{\alpha} \to \infty}\gamma_{F,\alpha}^{\pm }(\mathbf{T}_{\alpha}) =\mp \mathbf{\Sigma}_{\alpha} \cdot \mathbf{x}_{\alpha}.
 \tag{7-7}
\end{equation}
These are the correct pseudo Dirac and chirality operators on each commutative $ AdS^{2} $.
It is obvious that the Dirac operator (7-7) is $ \Lambda- $pseudo $ \dagger $-invariant:
\begin{equation}
D_{F,\alpha}^{(\pm )^{\dagger}}(\mathbf{T}_{\alpha})= \Lambda D_{F,\alpha}^{(\pm )}(\mathbf{T}_{\alpha})\Lambda^{-1},
\tag{7-8}
\end{equation}
which we expect from commutative Dirac operator in instanton sector on each $ AdS^{2} $.

Now, let us construct pseudo fuzzy Ginsparg-Wilson algebra for $ AdS_{F}^{2}\times AdS_{F}^{2} $ in instanton sector. It is a $ \Lambda- $pseudo $ \dagger- $invariant algebra over $ \mathbb{C} $,
\begin{equation}
\mathcal{A}_{l_{1},l_{2}}(\mathbf{T})= \langle \Gamma(\mathbf{T})= \Gamma_{1}(\mathbf{T}_{1})\Gamma_{2}(\mathbf{T}_{2}),\quad \Gamma^{'}= \Gamma_{1}^{'}\Gamma_{2}^{'}, \quad \Gamma^{2}= \Gamma^{'2}= 1, \quad \Gamma^{\dagger}= \Lambda\Gamma\Lambda^{-1}, \quad \Gamma^{'\dagger}= \Lambda\Gamma^{'}\Lambda^{-1} \rangle.\tag{7-9}
\end{equation}
where $ \Gamma_{1,2} $ and $ \Gamma_{1,2}^{'} $ are the generators of the Ginsparg-Wilson algebra of each $ AdS_{F}^{2} $ in instanton sector,which are given in (7-4).
Let us define the $ \Lambda- $pseudo fuzzy gauged Dirac operator on $ AdS_{F}^{2}\times AdS_{F}^{2} $ in instanton sector as:
\begin{equation}
D_{F}(\mathbf{T})= \dfrac{\Gamma_{1}(\mathbf{T}_{1})\Gamma_{2}(\mathbf{T}_{2})-\Gamma_{1}^{'}\Gamma_{2}^{'}}{2\sqrt{(l_{1}\pm t_{1})(1-l_{1}\mp t_{1})-1}\sqrt{(l_{2}\pm t_{2})(1-l_{2}\mp t_{2})-1}•}, \quad D_{F}(\mathbf{T})^{\dagger}= \Lambda D_{F}(\mathbf{T})\Lambda^{-1}.\tag{7-10}
\end{equation}
It is easy to see that
\begin{equation*}
\Gamma_{1}(\mathbf{T}_{1})\Gamma_{2}(\mathbf{T}_{2})-\Gamma_{1}^{'}\Gamma_{2}^{'}=
\end{equation*}
\begin{equation}
 \dfrac{1}{2•}[(\Gamma_{1}(\mathbf{T}_{1})- \Gamma_{1}^{'})(\Gamma_{2}(\mathbf{T}_{2}) +\Gamma_{2}^{'})+ (\Gamma_{1}(\mathbf{T}_{2})+\Gamma_{1}^{'})(\Gamma_{2}(\mathbf{T}_{2})- \Gamma_{2}^{'})]\tag{7-11}
\end{equation}
Now, using the following definitions
\begin{equation}
D_{1}^{F}(\mathbf{T}_{\alpha})= \dfrac{ (\Gamma_{1}(\mathbf{T}_{1})- \Gamma_{1}^{'})(\Gamma_{2}(\mathbf{T}_{2}) +\Gamma_{2}^{'}) }{2•\sqrt{(l_{1}\pm t_{1})(1-l_{1}\mp t_{1})-1}\sqrt{(l_{2}\pm t_{2})(1-l_{2}\mp t_{2})-1}}, \tag{7-12}
\end{equation}
\begin{equation}
 D_{2}^{F}(\mathbf{T}_{\alpha})= \dfrac{ (\Gamma_{1}(\mathbf{T}_{1})+ \Gamma_{1}^{'})(\Gamma_{2}(\mathbf{T}_{2}) -\Gamma_{2}^{'}) }{2•\sqrt{(l_{1}\pm t_{1})(1-l_{1}\mp t_{1})-1}\sqrt{(l_{2}\pm t_{2})(l_{2}\pm t_{2}+1)+1}}\tag{7-13}
\end{equation}
which satisfies
\begin{equation}
[D_{1}^{F}(\mathbf{T}_{\alpha}), D_{2}^{F}(\mathbf{T}_{\alpha})]=0\tag{7-14}
\end{equation}
the fuzzy Dirac operator (7-10) on $ AdS_{F}^{2}\times AdS_{F}^{2} $ in instanton sector can be written as:
\begin{equation}
D_{F}(\mathbf{T})= D_{1}^{F}(\mathbf{T}_{\alpha})+D_{2}^{F}(\mathbf{T}_{\alpha}).\tag{7-15}
\end{equation}
in the commutative limit (7-15) tends to 
\begin{equation*}
\lim_{l_{1,2} \to \infty }D_{F}(\mathbf{T})= D_{1}(\mathbf{T}_{\alpha})+D_{2}(\mathbf{T}_{\alpha})=
\end{equation*}
\begin{equation}
 (\mathbf{\Sigma}_{1}\cdot (\mathcal{L}_{1}+\mathbf{T_{1}})-1)(\mathbf{\Sigma}_{2}\cdot \mathbf{x}_{i_{2}})+(\mathbf{\Sigma}_{1}\cdot \mathbf{x}_{i_{1}})(\mathbf{\Sigma}_{2}\cdot (\mathcal{L}_{2}+\mathbf{T}_{2})-1).\tag{7-16}
\end{equation}

\section{Gauging the pseudo fuzzy Dirac operator in instanton sector}
The derivation $\mathcal{L}_{i_{1},i_{2}} $ dose not commute with the projectors $ P_{F}^{(l_{\alpha}\pm t_{\alpha})} $ and then has no action on the modules $Mat(2l_{\alpha}+1)P_{F}^{(l_{\alpha} \pm t_{\alpha})}$. But $J_{i_{\alpha}}=\mathcal{L}_{i_{\alpha}}+T_{i_{\alpha}}$ does commute with $ P_{F}^{(l_{\alpha}\pm t_{\alpha})}$ on each $ AdS_{F}^{2} $. Here, $ J_{i_{\alpha}} $ has been considered as the total angular momentum on each $ AdS_{F}^{2} $. Now, we need to gauge $J_{i_{\alpha}} $. When $ T=0$, the gauge fields $ A_{\alpha} $ were function of $ L_{\alpha}^{L}$. Here, we consider $ A_{\alpha}^{L} $ to be a functions of $ \mathbf{L}_{\alpha}^{L}+\mathbf{T}_{\alpha} $, because $ A_{\alpha}^{L} $ dose not commute with $ P_{F}^{(l_{\alpha}\pm t_{\alpha})} $. Let us introduce the covariant derivative as:
\begin{equation}
\nabla_{\alpha}=J_{\alpha}+A_{\alpha}^{L}.
\tag{8-1}
\end{equation}
In this case the limiting transversality of $ \mathbf{L}_{\alpha}^{L}+\mathbf{T}_{\alpha} $ can be guaranteed by imposing the condition:
\begin{equation}
(\mathbf{L}_{\alpha}^{L}+\mathbf{A}_{\alpha}^{L}+\mathbf{T}_{\alpha})\cdot(\mathbf{L}_{\alpha}^{L}+\mathbf{A}_{\alpha}^{L}+\mathbf{T}_{\alpha})=(\mathbf{L}_{\alpha}^{L}+\mathbf{T}_{\alpha})\cdot(\mathbf{L}_{\alpha}^{L}+\mathbf{T}_{\alpha})= (l_{\alpha}\pm t_{\alpha})(1-l_{\alpha}\mp t_{\alpha}),
\tag{8-2}
\end{equation}
The expansion of (8-2) is:
\begin{equation}
(\mathbf{L}_{\alpha}^{L}+\mathbf{T}_{\alpha})\cdot \mathbf{A}_{\alpha}^{L} +\mathbf{A}_{\alpha}^{L}\cdot(\mathbf{L}_{\alpha}^{L}+\mathbf{T}_{\alpha})+ \mathbf{A}_{\alpha}\cdot \mathbf{A}_{\alpha}=0.
\tag{8-3}
\end{equation}
When the parameter $ l_{\alpha} $ tends to infinity, $ \dfrac{A_{i_{\alpha}}^{L}}{l_{\alpha}} $ and $ \dfrac{\mathbf{T}_{\alpha}}{l_{\alpha}•} $ tend to zero and $ (L_{i_{1}}^{L}, L_{i_{2}}^{L}) $ tend to $ (x_{i_{1}}, x_{i_{2}}) $. Then, for large $ l_{\alpha} $, the (8-3) tends to the condition $ \mathbf{x}_{\alpha}\cdot \mathbf{a}_{\alpha}=0 $ on each $ AdS^{2} $.
Now, we can construct the gauged pseudo fuzzy Ginsparg-Wilson system in instanton sector and its corresponding fuzzy Dirac and chirality operators on each $ AdS_{F}^{2} $ as follow: 
\begin{equation*}
\mathcal{A}_{\alpha}^{\pm}( \mathbf{T}_{\alpha},\mathbf{A}_{\alpha}^{L})=
\end{equation*}
\begin{equation}
 \langle \Gamma_{\alpha}^{\pm}(\mathbf{T_{\alpha},A_{\alpha}^{L}}), \Gamma_{\alpha}^{'} : \Gamma_{\alpha}^{\pm^{^{2}}}(\mathbf{T_{\alpha},A_{\alpha}^{L}})=\Gamma_{\alpha}^{'^{2}}=1,\quad\Gamma_{\alpha}^{\pm \dagger}(\mathbf{T_{\alpha},A_{\alpha}^{L}})=\Lambda\Gamma_{\alpha}^{\pm}(\mathbf{T}_{\alpha})\Lambda^{-1} ,\quad \Gamma_{\alpha}^{'\dagger}=\Lambda\Gamma_{\alpha}^{'}\Lambda^{-1} \rangle.
\tag{8-4}
\end{equation}
We introduce the involutory $ \Lambda- $pseudo $ \dagger $-invariant generators of the Ginsparg-Wilson system as:
\begin{equation}
\Gamma_{\alpha}^{\pm }(\mathbf{T}_{\alpha},\mathbf{A}_{\alpha}^{L}) = \mp \dfrac{\mathbf{\Sigma }_{\alpha}\cdot(\mathbf{L}_{\alpha}^{L}+ \mathbf{T}_{\alpha}+\mathbf{A}_{\alpha}^{L})-1}{|\mathbf{\Sigma }_{\alpha}\cdot(\mathbf{L}_{\alpha}^{L}+ \mathbf{T}_{\alpha}+\mathbf{A}_{\alpha}^{L})-1|}\:,\qquad \Gamma_{\alpha}^{' \pm }= \pm \dfrac{\mathbf{\Sigma }_{\alpha}\cdot \mathbf{L}_{\alpha}^{R}+1}{|\mathbf{\Sigma }_{\alpha}\cdot(\mathbf{L}_{\alpha}^{R})+1|} .
\tag{8-5}
\end{equation}
Now, up to the first order (8-5) becomes:
\begin{equation}
\Gamma_{\alpha}^{\pm }(\mathbf{T}_{\alpha},\mathbf{A}_{\alpha}^{L}) = \mp \dfrac{\mathbf{\Sigma }_{\alpha}\cdot(\mathbf{L}_{\alpha}^{L}+ \mathbf{T}_{\alpha}+\mathbf{A}_{\alpha}^{L})-1}{\sqrt{(l_{\alpha}\pm t_{\alpha})(1-l_{\alpha}\mp t_{\alpha})-1}} \:,\qquad \Gamma_{\alpha}^{' \pm }= \pm \dfrac{\mathbf{\Sigma }_{\alpha}\cdot \mathbf{L}_{\alpha}^{R}+1}{\sqrt{l_{\alpha}(1-l_{\alpha})-1}} .
\tag{8-6}
\end{equation}
Using the definitions (5-2), (5-5) and (8-6) one can calculate $ \Lambda- $pseudo fuzzy Dirac and chirality operators on each $ AdS_{F}^{2} $, which in the commutative limit they become:
\begin{equation}
 \lim_{l_{\alpha} \to \infty} D_{F,\alpha}^{\pm }(\mathbf{T}_{\alpha}, \mathbf{A}_{\alpha}^{L})=\pm(\mathbf{\Sigma}_{\alpha} \cdot (\mathcal{L}_{\alpha}+ \mathbf{T}_{\alpha}+ \mathbf{A}_{\alpha}^{L})-1),\qquad  \lim_{l_{\alpha} \to \infty}\gamma_{F,\alpha}^{\pm }(\mathbf{T}_{\alpha}, \mathbf{A}_{\alpha}^{L}) =\pm \mathbf{\Sigma}_{\alpha} \cdot \mathbf{x}_{\alpha},
 \tag{8-7}
\end{equation}
which we expect from commutative gauged Dirac and chirality operators in instanton sector.

Now, let us construct gauged pseudo fuzzy Ginsparg-Wilson algebra for $ AdS_{F}^{2}\times AdS_{F}^{2} $ in instanton sector. It is a $ \Lambda- $pseudo $ \dagger- $invariant algebra over $ \mathbb{C} $,
\begin{equation*}
\mathcal{A}_{l_{1},l_{2}}(\mathbf{A}^{L}, \mathbf{T})= 
\end{equation*}
\begin{equation}
\langle \Gamma(\mathbf{A}^{L},\mathbf{T})= \Gamma_{1}(\mathbf{A}_{1}^{L},\mathbf{T}_{1})\Gamma_{2}(\mathbf{A}_{2}^{L},\mathbf{T}_{2}),\quad \Gamma^{'}= \Gamma_{1}^{'}\Gamma_{2}^{'}, \quad \Gamma^{2}= \Gamma^{'2}= 1, \quad \Gamma^{\dagger}= \Lambda\Gamma\Lambda^{-1}, \quad \Gamma^{'\dagger}= \Lambda\Gamma^{'}\Lambda^{-1} \rangle.\tag{8-8}
\end{equation}
where $ \Gamma_{1,2} $ and $ \Gamma_{1,2}^{'} $ are the generators of the gauged Ginsparg-Wilson algebra of each $ AdS_{F}^{2} $ in instanton sector,which are given in (8-6).
Let us define the fuzzy $ \Lambda- $pseudo gauged Dirac operator on $ AdS_{F}^{2}\times AdS_{F}^{2} $ in instanton sector as:
\begin{equation}
D_{F}(\mathbf{A}^{L},\mathbf{T})= \dfrac{\Gamma_{1}(\mathbf{A}_{1}^{L},\mathbf{T}_{1})\Gamma_{2}(\mathbf{A}_{2}^{L},\mathbf{T}2)-\Gamma_{1}^{'}\Gamma_{2}^{'}}{2\sqrt{(l_{1}\pm t_{1})(1-l_{1}\mp t_{1})-1} \sqrt{(l_{2}\pm t_{2})(1-l_{2}\mp t_{2})-1}•}, \quad  D_{F}(\mathbf{A}^{L},\mathbf{T})^{\dagger}= \Lambda D_{F}(\mathbf{A}^{L},\mathbf{T})\Lambda^{-1}\tag{8-9}
\end{equation}
It is easy to see that
\begin{equation*}
\Gamma_{1}(\mathbf{A}_{1}^{L},\mathbf{T}_{1})\Gamma_{2}(\mathbf{A}_{2}^{L},\mathbf{T}_{2})-\Gamma_{1}^{'}\Gamma_{2}^{'}=
\end{equation*}
\begin{equation}
 \dfrac{1}{2•}[(\Gamma_{1}(\mathbf{A}_{1}^{L},\mathbf{T}_{1})- \Gamma_{1}^{'})(\Gamma_{2}(\mathbf{A}_{2}^{L},\mathbf{T}_{2}) +\Gamma_{2}^{'})+ (\Gamma_{1}(\mathbf{A}_{1}^{L},\mathbf{T}_{2})+\Gamma_{1}^{'})(\Gamma_{2}(\mathbf{A}_{2}^{L},\mathbf{T}_{2})- \Gamma_{2}^{'})]\tag{8-10}
\end{equation}
Now, using the following definitions
\begin{equation}
D_{1}^{F}(\mathbf{A}_{\alpha}^{L}, \mathbf{T}_{\alpha})= \dfrac{ (\Gamma_{1}(\mathbf{A}_{1}^{L}, \mathbf{T}_{1})- \Gamma_{1}^{'})(\Gamma_{2}(\mathbf{A}_{2}^{L}, \mathbf{T}_{2}) +\Gamma_{2}^{'}) }{2•\sqrt{(l_{1}\pm t_{1})(1-l_{1}\mp t_{1})-1}\sqrt{(l_{2}\pm t_{2})(1-l_{2}\pm t_{2})-1}}, \tag{8-11}
\end{equation}
\begin{equation}
 D_{2}^{F}(\mathbf{A}_{\alpha}^{L}, \mathbf{T}_{\alpha})= \dfrac{ (\Gamma_{1}(\mathbf{A}_{1}^{L},\mathbf{T}_{1})+ \Gamma_{1}^{'})(\Gamma_{2}(\mathbf{A}_{2}^{L},\mathbf{T}_{2}) -\Gamma_{2}^{'}) }{2•\sqrt{(l_{1}\pm t_{1})(1-l_{1}\pm t_{1})-1}\sqrt{(l_{2}\pm t_{2})(1-l_{2}\pm t_{2})-1}}\tag{8-12}
\end{equation}
which satisfies
\begin{equation}
[D_{1}^{F}(\mathbf{A}_{\alpha}^{L}, \mathbf{T}_{\alpha}), D_{2}^{F}(\mathbf{A}_{\alpha}^{L}, \mathbf{T}_{\alpha})]=0,\tag{8-13}
\end{equation}
the fuzzy Dirac operator (8-9) on $ AdS_{F}^{2}\times AdS_{F}^{2} $ in instanton sector can be written as:
\begin{equation}
D_{F}(\mathbf{A}^{L}, \mathbf{T})= D_{1}^{F}(\mathbf{A}_{\alpha}^{L}, \mathbf{T}_{\alpha})+D_{2}^{F}(\mathbf{A}_{\alpha}^{L}, \mathbf{T}_{\alpha}).\tag{8-14}
\end{equation}
in the commutative limit (8-14) tends to 
\begin{equation*}
\lim_{l_{1,2} \to \infty }D_{F}(\mathbf{A}^{L}, \mathbf{T})= D_{1}(\mathbf{A}_{\alpha}, \mathbf{T}_{\alpha})+D_{2}(\mathbf{A}_{\alpha}, \mathbf{T}_{\alpha})=
\end{equation*}
\begin{equation}
 (\mathbf{\Sigma}_{1}\cdot (\mathcal{L}_{1}+ \mathbf{A}_{1}+\mathbf{T_{1}})-1)(\mathbf{\Sigma}_{2}\cdot \mathbf{x}_{i_{2}})+(\mathbf{\Sigma}_{1}\cdot \mathbf{x}_{i_{1}})(\mathbf{\Sigma}_{2}\cdot (\mathcal{L}_{2}+\mathbf{A}_{2}+\mathbf{T}_{2})-1).\tag{8-15}
\end{equation}

\section{Conclusion}
\textbf{}
In this paper, using the $ \Lambda- $pseudo projectors and idempotents of the finitely generated projective $ A(AdS_{F}^{2}\times AdS_{F^{2}})  -$module of the principal fibration $ X_{AdS_{F}}^{5}\rightarrow  AdS_{F}^{2}\times AdS_{F}^{2}$ it has been constructed the generators of the gauged fuzzy Ginsparg-Wilson algebra in instanton sector. It has been constructed gauged fuzzy Dirac operator in instanton sector using the fuzzy Ginsparg-Wilson algebra. The importance of this Dirac operator is that it has correct commutative limit.

\end{document}